\documentclass[manuscript]{aastex}
\usepackage{color}

\shorttitle{Rotation}
\shortauthors{Iwakami et al.}

\begin{document}


\title{Critical Surface for Explosions of Rotational Core-Collapse Supernovae
      }


\author{Wakana Iwakami}
\affil{
Yukawa Institute for Theoretical Physics, Kyoto University, Oiwake-cho, Kitashirakawa, Sakyo-ku, Kyoto, 606-8502, Japan}
\affil{Advanced Research Institute for Science and Engineering, Waseda University, 3-4-1, Okubo, Shinjuku, Tokyo, 169-8555, Japan}

\author{Hiroki Nagakura}
\affil{
Yukawa Institute for Theoretical Physics, Kyoto University, Oiwake-cho, Kitashirakawa, Sakyo-ku, Kyoto, 606-8502, Japan}

\and

\author{Shoichi Yamada}
\affil{Advanced Research Institute for Science and Engineering, Waseda University, 3-4-1, Okubo, Shinjuku, Tokyo, 169-8555, Japan}
\email{wakana@heap.phys.waseda.ac.jp}

\begin{abstract}
The effect of rotation on the explosion of core-collapse supernovae is investigated systematically in three-dimensional simulations.
In order to obtain the critical conditions for explosion as a function of mass accretion rate, neutrino luminosity, and specific angular momentum, rigidly rotating matter was injected from the outer boundary with an angular momentum, which is increased every 500 ms.
It is found that there is a critical value of the specific angular momentum, above which the standing shock wave revives, for a given combination of mass accretion rate and neutrino luminosity, i.e. an explosion can occur by rotation even if the neutrino luminosity is lower than the critical value for a given mass accretion rate in non-rotational models.
The coupling of rotation and hydrodynamical instabilities plays an important role to characterize the dynamics of shock revival for the range of specific angular momentum that are supposed to be realistic.
Contrary to expectations from past studies, the most rapidly expanding direction of the shock wave is not aligned with the rotation axis.
Being perpendicular to the rotation axis on average, it can be oriented in various directions.
Its dispersion is small when the spiral mode of the standing accretion shock instability (SASI) governs the dynamics, while it is large when neutrino-driven convection is dominant.
As a result of the comparison between 2D and 3D rotational models, it is found that $m\ne 0$ modes of 
neutrino-driven convection
or SASI are important for shock revival around the critical surface.
\end{abstract}


\keywords{supernovae --- hydrodynamics --- instabilities --- rotation}



\section{INTRODUCTION \label{sec_intro}}

A number of observational results and numerical studies on core-collapse supernovae have indicated that multi-dimensional effects are important in their explosions.
Observations have revealed that explosions are asymmetric \citep[e.g.][]{wang01, leonard06}, and one-dimensional (1D) simulations of spherically symmetric core collapse have shown that explosions do not obtain, except for low mass progenitors \citep[e.g.][]{liebend05,sumiyoshi05}.
Hence, a large number of multi-dimensional numerical studies have been done to investigate 
the influences of various (magneto) hydrodynamical instabilities,
rotation, and magnetic fields on the explosion \citep[see][for latest reviews]{janka12, burrows13}.
In this paper, we focus on rotation, particularly its effects on shock revival.
We now know that there is a critical neutrino luminosity for a given mass accretion rate, above which a stalled shock wave is revived \citep{burrows93}.
We are interested in the change that rotation will make in the critical luminosity.

In multi-dimensional simulations of the post bounce phase in core-collapse supernovae, two types of hydrodynamical instabilities, i.e.
neutrino-driven convection and standing accretion shock instability (SASI), are commonly observed before shock revival \citep[e.g.][]{muller12, ott13, hanke13, couch13b}.
To be more specific, by the neutrino-driven convection we mean in this paper the hydrodynamical instability caused by the negative entropy gradient
that is produced in the gain region via heating of matter by neutrinos
radiated from a proto-neutron star \citep[e.g.][]{bethe90, herant94, janka96}.
The SASI, on the other hand, stands for the instability induced by the advective-acoustic cycle and accompanied by sloshing and/or spiral motions of the shock front \citep[e.g.,][]{foglizzo07, yamasaki07, yamasaki08, foglizzo09, fernandez09, guilet12}.
There are some two-dimensional (2D) and three-dimensional (3D) simulations that demonstrated the dominance of neutrino-driven convection \citep[e.g.,][]{fryer02,nordhaus10,burrows12,couch13a,dolence13, murphy13}.
The SASI has been also confirmed by 2D and 3D simulations
\citep[e.g.,][]{blondin03, ohnishi06, blondin07a, blondin07b, iwakami08a, scheck08, fernandez10, takiwaki12, muller12, hanke12, hanke13, ott13} as well as by
SWASI experiment \citep{foglizzo12}.
In the nonlinear regime, SASI tends to produce coherent large-scale structures \citep[e.g.][]{blondin07a} whereas the neutrino-driven convection induces turbulence \citep[e.g.][]{murphy13}. 
It has been also demonstrated numerically that spatial dimension is another key parameter to determine the characteristics of these instabilities \citep[e.g.][]{burrows12}.
There is an ongoing dispute on which instability plays more important roles in shock revival.

Recently, we have published a parametric study by 3D simulations of flow patterns produced by the
neutrino-driven convection 
and SASI \citep{iwakami14}.
In fact a particular type of flow pattern emerges according to the mass accretion rate and neutrino luminosity.
Buoyant bubbles tend to be formed for the combination of low mass accretion rates and high neutrino luminosities whereas spiral or sloshing pattern appears as the mass accretion rate increases and/or the neutrino luminosity decreases.
These patterns are supposed to be associated with the growths of 
neutrino-driven convection and SASI.
It is first shown by linear analysis \citep{foglizzo06} and then confirmed by numerical simulations \citep[e.g.][]{scheck08} that the neutrino-driven convection occurs when the ratio of advection time to linear growth time, or the $\chi$ parameter, is larger than $\sim$ 3 and SASI appears otherwise.
We found that the bubble formation occurs for $\chi \gtrsim 3$ whereas the spiral or sloshing pattern appears otherwise.
This seems to support the claim that the bubble formation is associated with the development of neutrino-driven convection while the sloshing and spiral patterns are consequences of SASI.
In that paper, rotation is entirely ignored.
We are hence interested in the influence of rotation on the pattern formation in the post-shock flow. 

Observational data clearly indicate that massive stars rotate rapidly on the main sequence \citep{tassoul78, maeder12}.
Possible effects of rotation on the core-collapse supernovae have been investigated by 2D axisymmetric simulations.
The rotationally induced non-spherical structure of the progenitor may explain strong matter mixing as observed in SN1987A
\citep[e.g.][]{chevalier89, yamada90, yamada91}.
It was found that rotation has a negative effect on the prompt explosion mechanism
\citep[e.g.][]{muller81, yamada94, fryer00},
since centrifugal force prevents the iron core from contracting sufficiently, and less gravitational energy becomes available.
In the context of the neutrino-heating mechanism, which is our concern in this paper, it was argued that the enhancement of neutrino heating near the rotation axis may induce a jet-like explosion \citep{kotake03}.
Linear analysis of steady accretion flows through a standing shock wave also suggested that the shock revival takes place at the rotation axis \citep{yamasaki05}.
More importantly, that study demonstrated that the critical luminosity can be lowered by rapid rotation.
On the other hand, some 2D realistic simulations found that the
rotation-induced anisotropy in neutrino-heating was not sufficient in the explosion \citep{buras03, walder05}.
3D numerical studies of rotational models have also been done:
\citet[][]{fryer04} presented the results of smoothed particle hydrodynamics (SPH) simulations; \citet[][]{iwakami09a, iwakami09b} investigated the influence of rotation on SASI;
very recently \citet{nakamura14} performed realistic simulations with detailed neutrino transfer.
Unfortunately, it is not clear in these studies how rotation affects the shock revival as well as the development of neutrino-driven convection and/or SASI.
In this paper, we investigate it in a systematic manner with 3D hydrodynamical simulations.
They are rather experimental, since we are mainly concerned with the systematics: only the post-bounce phase is considered; the central region inside the neutrino sphere is excised and neutrino transport is replaced by the light-bulb approximation.

The organization of this paper is as follows.
Models and numerical setups are described in Section 2, results are presented in Section 3, and conclusion is given in Section 4.

\section{MODELS AND NUMERICAL SETUPS \label{sec_models}}

An explosion of a core-collapse supernova has a lot of stages.
In this paper, we focus only on the shock revival after the core bounce. 
A computational domain covers the inner part of the iron core, where the spherical coordinate system $(r, \theta, \phi)$ is considered.
The central part, corresponding to the proto-neutron star (PNS), is excised from the computational domain.
The initial flows are the 1D steady solutions 
\citep{yamasaki06}.
The matter is accreted from the outer boundary onto the PNS with a fixed mass accretion rate,
and the spherical shock wave is stalled on its way.
The neutrino sphere, being assumed to be inside the PNS, isotropically gives off neutrinos with a fixed neutrino flux to the optically thin matter.
The neutrino radiation makes a negative entropy gradient in the gain region around the PNS.
More details about the initial and boundary conditions are described in our previous paper for non-rotational models \citep{iwakami14}.
The amplitude of the initial perturbation for the radial velocity $\delta{v_r}/v_r$ is determined at random in every cell within 1\%.

The basic equations are the three-dimensional compressible Euler equations
\citep[][]{iwakami08a}.
Thermodynamical variables are obtained from Shen's equation of state (EOS) \citep{shen98}.
The neutrino heating/cooling term is calculated with the light-bulb approximation \citep{ohnishi06}.
These equations are numerically solved by ZEUS-MP/2 code \citep{hayes06}.
A tensor-type artificial viscosity is used for suppressing the carbuncle phenomenon around the polar axis \citep{iwakami08a, iwakami08b}.
The computational domain
covers the region on the entire solid angle from $r_{\rm in}$ to $r_{\rm out}$,
where $r_{\rm in}$ and $r_{\rm out}$ are the radii of the inner and outer boundaries.
It is divided into $N_r\times N_\theta \times N_\phi = 300\times30\times60$ grid cells,
where $N_r$, $N_\theta$, and $N_\phi$ denote the numbers of grid cells in $r$, $\theta$, and $\phi$ directions, respectively.
The radial grid width is determined to be 1\% of the radius everywhere, and the polar and azimuthal grids are uniformly spaced to avoid numerical oscillations around the polar axis.
Results of test runs with higher angular resolutions are summarized in Appendix \ref{appex_res}.
The temperatures of electron-type neutrinos and anti-neutrinos are set to be the typical values in the post-bounce phase, $T_{\nu_{\mathrm{e}}} = 4 $~MeV and $T_{\bar{\nu}_{\mathrm{e}}} = 5 $~MeV, respectively.
The mass of the PNS is fixed to be $M_{\rm PNS} =1.4 M_\odot$.
The range of neutrino luminosities and mass accretion rates are $L_\nu = 2.5 - 6.0 \times 10^{52}$~erg~s$^{-1}$ and $\dot{M} = 0.2 - 1.0~M_{\odot}$~s$^{-1}$, respectively.
These parameters of all models are summarized in Table~\ref{tbl_param}.

In this study, rigidly rotating matter is injected from the outer boundary surface into the computational region with an angular momentum increased every 500 ms.
The azimuthal velocity imposed on the outer boundary is written in
\begin{equation}
v_\phi(r_{out},\theta) = \beta \times  10^{8} \times \left(\frac{10^8}{r_{out}}\right) \times \sin \theta \ [{\rm cm\  s}^{-1}],
\label{eq_rot}
\end{equation}
where $\beta$ is a parameter of rotation.
The detailed derivation of Eq.~(\ref{eq_rot}) and its validity can be referred to in Appendix \ref{appex_rot}.
The corresponding specific equatorial angular momentum is described as
\begin{equation}
L = \beta \times  10^{16} \  [{\rm cm}^{2}\ {\rm s}^{-1}].
\label{eq_san}
\end{equation}
According to \citet{heger00, heger05}, the 15 $M_\odot$ pre-SN star has
$L \sim 10^{14} - 10^{15}$ cm$^2$ s$^{-1}$ with magnetic fields
and $L \sim 10^{16} - 10^{17}$ cm$^2$ s$^{-1}$ without ones, in the range of the enclosed mass from $\sim 1.4 M_\odot$ to $\sim 2.5 M_\odot$.
In this paper, the range of $\beta$ is taken from 0.0 to 1.0,
and we call $\beta$ the normalized specific angular momentum.

\section{RESULTS AND DISCUSSIONS \label{sec_result}}

\subsection{Critical Surface for Explosion \label{sec_crit}}
\begin{figure}
\epsscale{.90}
\plotone{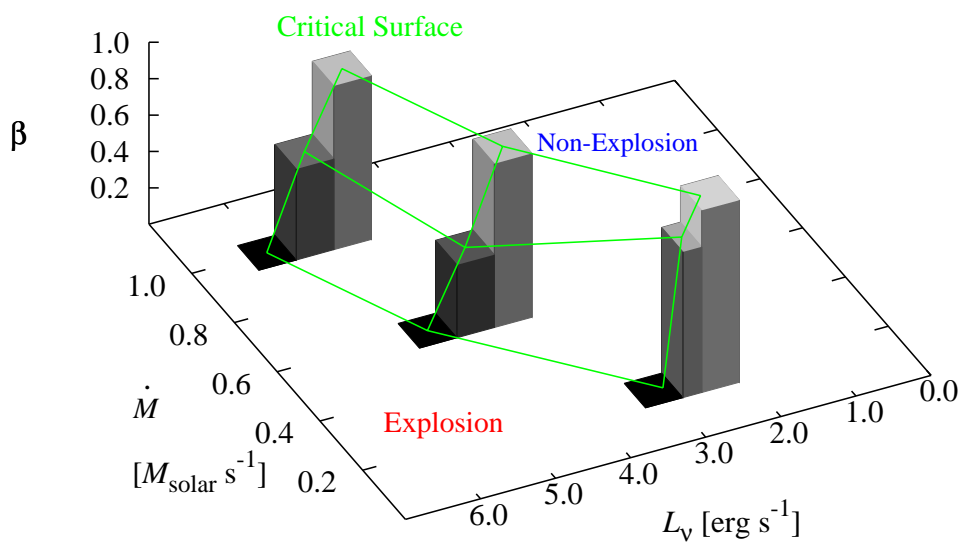}
\caption{
The critical surface for explosion as functions of the mass accretion rate $\dot{M}$, the neutrino luminosity $L_\nu$, and the normalized specific angular momentum $\beta$.
The critical surface is depicted with green lines.
The front side of the critical surface corresponds to the explosion region, and the other side agrees with the non-explosion region. 
The hight and color of vertical bars indicate the critical values of the normalized specific angular momentum $\beta_\mathrm{exp}$. 
}
\label{fig_crit}
\end{figure}

\citet{burrows93} discovered that there is a critical neutrino luminosity $L_{\nu}$, above which there is no steady solution, for a given mass accretion rate $\dot{M}$.
A critical curve, dividing the parameter space into explosion and non-explosion areas, can be depicted in the $L_{\nu}-\dot{M}$ plane.
In this study, we add one more parameter, a normalized specific angular momentum $\beta$, to $L_{\nu}$ and $\dot{M}$.
Then we draw a critical surface, above which the shock wave keep propagating outward at least up to 500 km, in the three-dimensional parameter space with $\dot{M}$, $L_{\nu}$ and $\beta$ axes in Figure~\ref{fig_crit}.
In order to obtain the critical surface, rigidly rotating matter is injected from the outer boundary into the supersonic inflow with its $\beta$ increased every 500 ms. 
We find that there is a critical value of $\beta$, represented by the monochrome boxes in Figure~\ref{fig_crit}, for a given combination of $\dot{M}$ and $L_\nu$.
The more rapidly rotating the matter is, the lower neutrino luminosity the shock revival requires.
The green lines, connected with the black squares on the plane at $\beta=0.0$, can be considered as the critical curve for non-rotational models, and the critical curve moves toward low $L_\nu$ as $\beta$ increases.
Therefore, it is confirmed that the rotation makes easy to explode in the 3D models.
Here, we notify that 
the critical curve at $\beta=0.0$ is located in somewhat higher $L_\nu$ than the curve given by more realistic simulations \citep{nordhaus10, hanke12, couch13a}, although it agrees well with the curve obtained from the analytical study, taking 
neutrino-driven convection 
into account, in the almost same setups as ours \citep{yamasaki06}.
Hence, we estimate that the critical surface, given by more sophisticated simulations, moves toward lower $L_\nu$ than the surface obtained in this study.
The effect of grid resolution on the critical surface is also discussed in Appendix \ref{appex_res}.

\begin{figure}
\epsscale{1.0}
\plotone{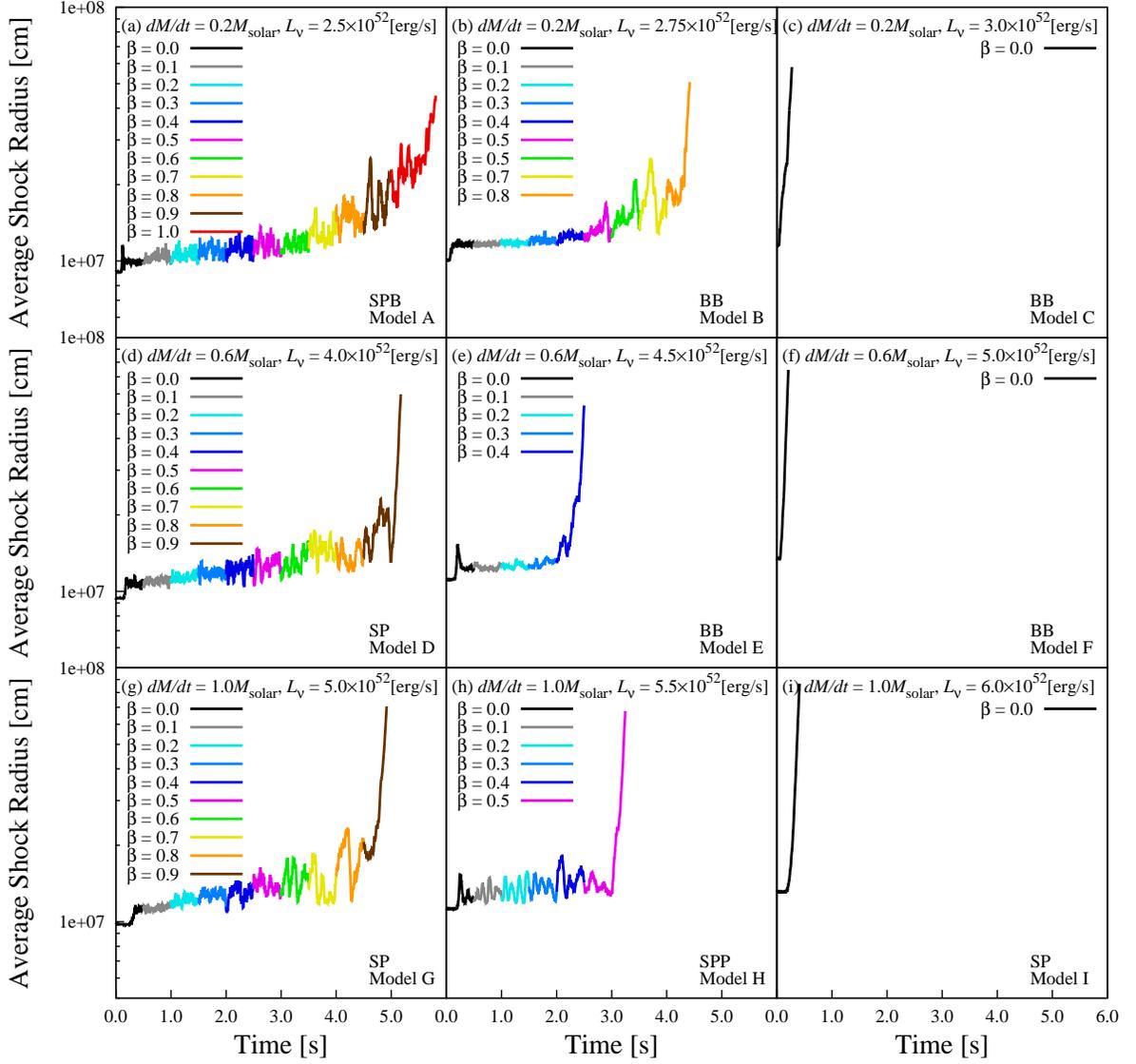}
\caption{
The evolutions of the averaged shock radius with increasing $\beta$ every 500 ms.
}
\label{fig_rsh}
\end{figure}

The time evolutions of the average shock radius are presented in Figure~\ref{fig_rsh}.
They are calculated with the expansion coefficient $c_0^0$ in Equation~(\ref{eq_c}) multiplied by $K_0^0$ in Equation~(\ref{eq_k}).
Rotation makes the shock radius enlarged, and more rapid rotation causes shock revival, even if the neutrino luminosity is not enough large to induce shock revival for non-rotational models.
In the slow-rotational models for $\beta \lesssim 0.4$, 
the evolution of the shock wave 
reflect the features of flow patterns which are developed behind the shock wave for non-rotational models.
In our previous study, the flow patterns are classified into sloshing motion (SL), spiral motion (SP), high-entropy bubbles (BB), spiral motion with rising buoyant bubbles (SPB), and spiral motion with pulsating rotational velocity (SPP).
The analysis of $\chi$ parameter indicates that SL and SP are caused by the growth of SASI, and BB arises as a result of 
neutrino-driven convection.
The remaining two patterns, SPB and SPP, are considered as intermediate patterns of SP and BB.
The abbreviated expressions of these patterns for non-rotational models are written in the bottom-right corners of the panels in Figure~\ref{fig_rsh}.
More detailed characteristics of these non-rotational models except Model B are described in our previous paper \citep{iwakami14}.
The properties of irrotational Model B are summarized in Appendix \ref{appex_B}.

Figure~\ref{fig_nht} shows the time evolutions of the net heating rate behind the shock wave.
It is clear that the time variations of them are correlated with those of the shock radii in Figure~\ref{fig_rsh}.
Therefore, the neutrino heating mechanism is considered to work predominantly for the shock revival at least for the range of $\beta$ from 0.0 to 1.0.
As the speed of rotation is amplified, the net heating rates tend to increase with growing the shock radii.
Hence, it is obvious that the enhancement of neutrino heating can be involved by rotation.

\begin{figure}
\epsscale{1.0}
\plotone{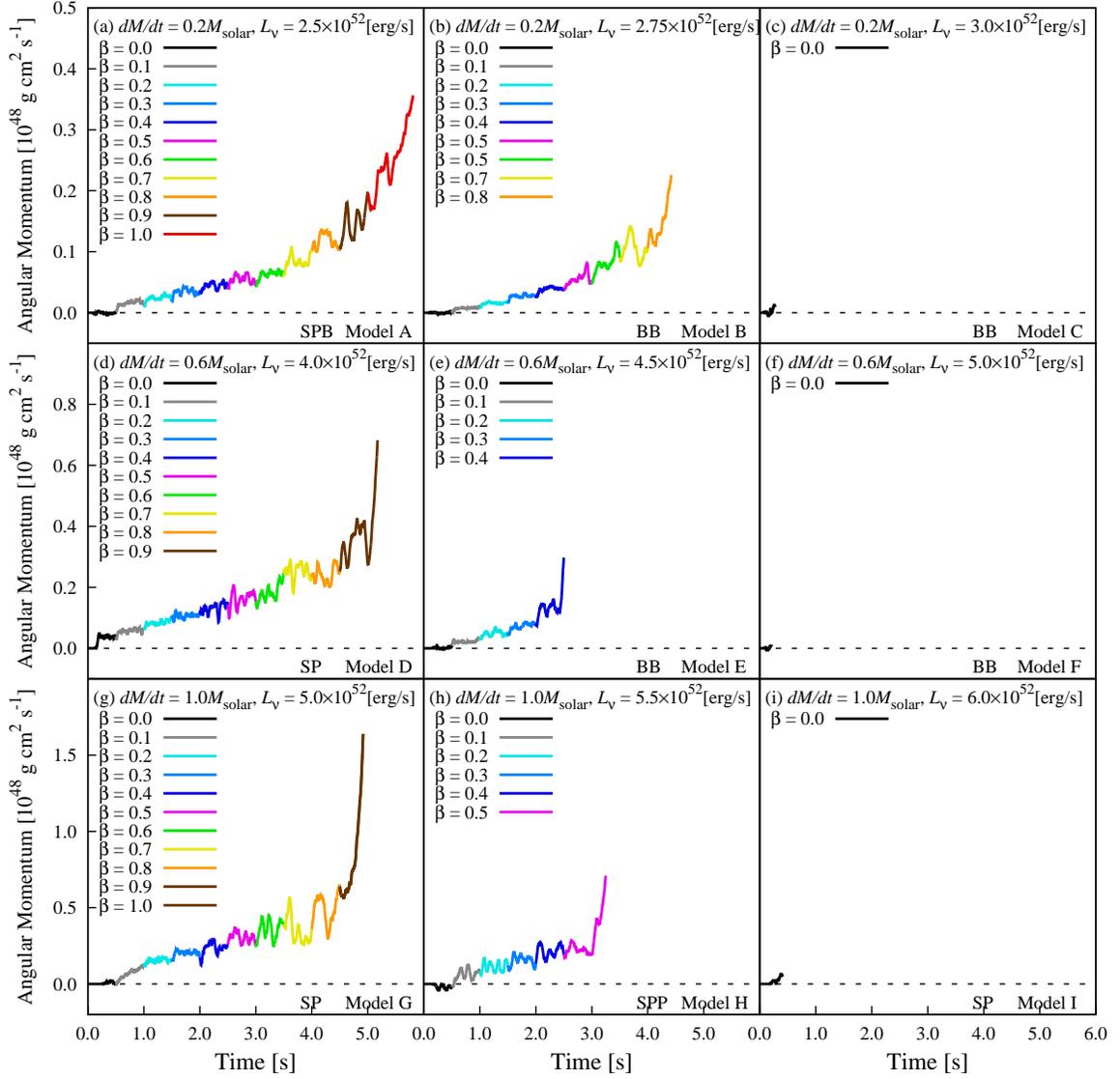}
\caption{
The evolutions of the angular momentum in the polar direction with increasing $\beta$ every 500 ms.
The angular momentum is integrated in the shocked flow. 
}
\label{fig_ang}
\end{figure}

The evolution of the angular momentum in the flow behind the shock wave is shown in Figure~\ref{fig_ang}.
The shocked matter has the angular momentum, which rotates in the same direction as the injected matter.
The magnitude of the angular momentum
is proportional to $\beta$ in the slow rotation phase,
and it exponentially increases before shock revival.
Such amplifications of the angular momentum can induce the strong centrifugal force to the shocked matter, and they are supposed to change the characteristics of flow dynamics and neutrino heating/cooling.

The two possibilities, how the shock revival can occur in the rotating cores, are considered here.
One is the prolate explosion: shock expansion begins from the rotation axis at the onset of the explosion \citep{yamasaki05}.
The other is the oblate explosion:
the preferred direction of explosion is perpendicular to the rotation axis \citep[][]{nakamura14}.
In the next section, we focus on the expanding direction of the shock wave at the shock revival.

\subsection{Expanding Direction of the Shock Wave at the Shock Revival \label{sec_direc}}

\begin{figure}
\epsscale{1.0}
\plotone{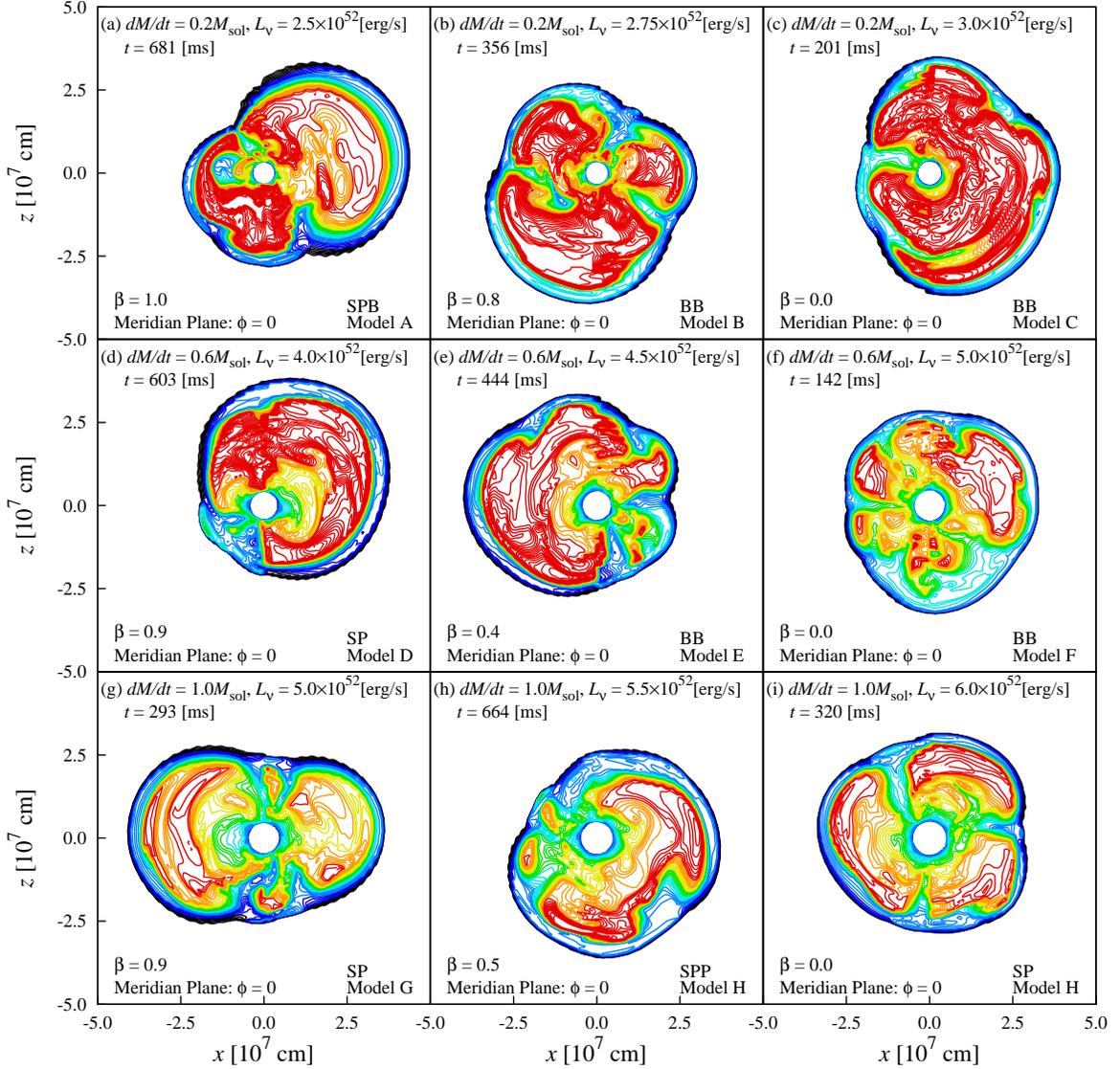}
\caption{
The snapshots of the entropy contour maps in the meridian plane at $\phi=0$ just after the shock revival.
The entropy $S$ is
in units of Boltzmann's constant $k_b$ per nucleon.
The contour levels are
in the range of $4 \le S \le 26$ with the increment of $\Delta S = 0.4$.
The contour lines of high values are drawn in reddish colors, and those of low ones are done in bluish colors.
The innermost and outermost contour lines agree with the surfaces of the proto-neutron star and the shock wave, respectively.
\label{fig_ent}}
\end{figure}

Figure~\ref{fig_ent} presents the snapshots of the entropy contour maps on the meridian plane at $\phi=0$ for explosion models, when the average shock radii reach 300 km after shock revival.
The left, middle, and right three panels show the entropy distributions for explosion models with rapid-rotation, slow-rotation, and non-rotation, respectively.
The shock wave geometry has a strong unipolar nature in the two rapid rotation models (Figure~\ref{fig_ent} (a) and (d)), and it has a dipolar nature in the other rapid rotation one (Figure~\ref{fig_ent} (g)).
On the other hand, 
such directional characteristics are not clear
in the non-rotational models,
since high entropy bubbles are formed in various directions
(Figure~\ref{fig_ent} (c), (f), (i)).
In the slow rotation models, the weak unipolar explosions can be observed (Figure~\ref{fig_ent} (b), (e), and (h)). 
Although several realizations should be done for each model
in order to confirm whether these directional properties are robust, it is obvious that the expanding directions of the shock waves are not specified parallel nor perpendicular to the rotation axis for the range of $\beta$ from 0.1 to 1.0.

\begin{figure}
\epsscale{1.0}
\plotone{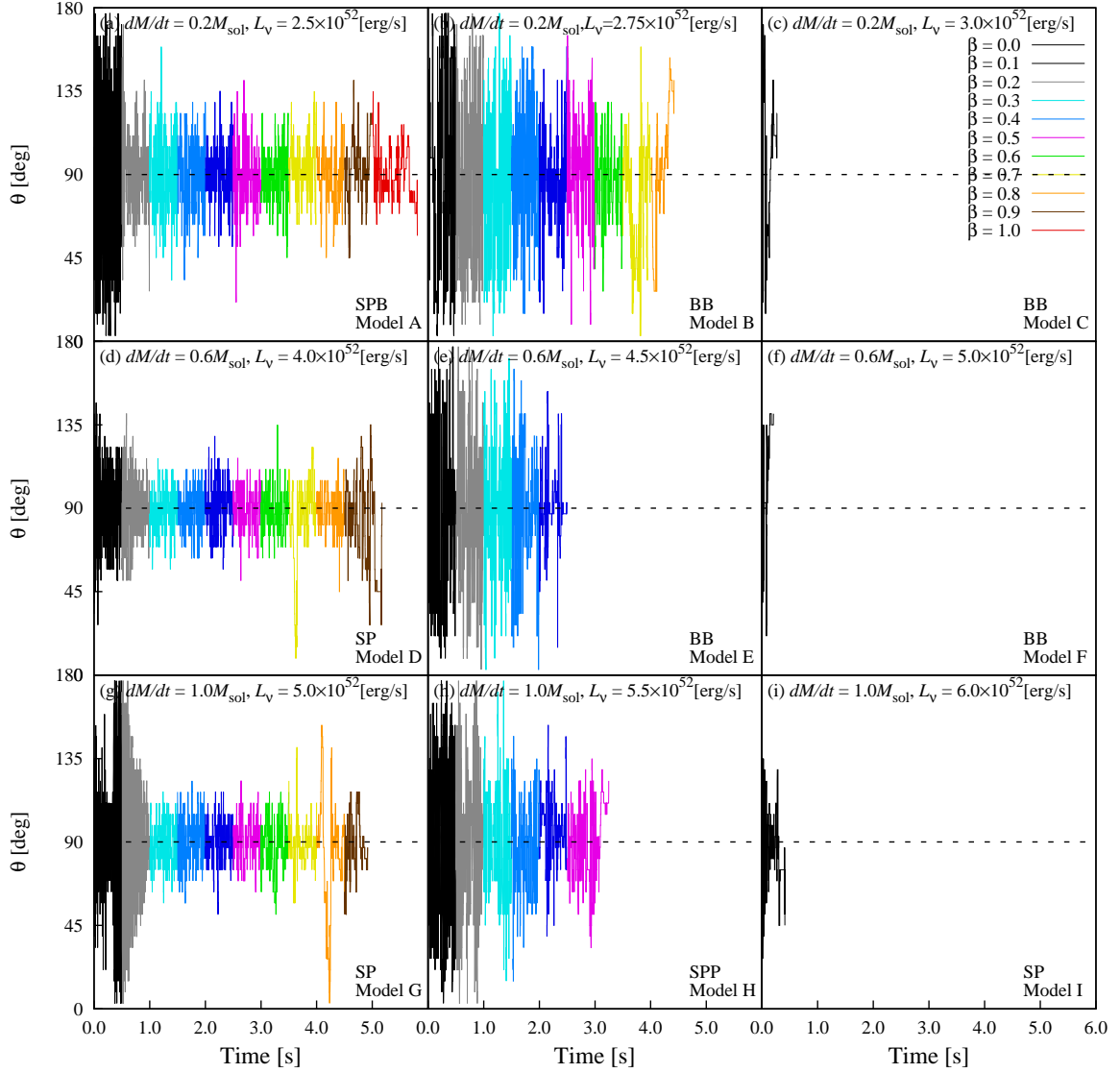}
\caption{
Time evolutions of the position angle, at which the shock wave is most extended, with increasing $\beta$ every 500 ms.
}
\label{fig_th}
\end{figure}

The evolutions of the position angle, at which the shock wave is most extended, are shown in Figure~\ref{fig_th}.
The black lines correspond to the results of non-rotational models.
The position angles $\theta$ are varied, depending on the dominant flow patterns.
Once a rotational flow passes through the shock wave, the rotation axis of the flow behind the shock wave is fixed to be parallel to the polar axis in all cases, and the direction, at which the shock radius is most extended, prefers to be perpendicular to the rotation axis due to the centrifugal force.
Especially, 
the variances of $\theta$ from 90$^\circ$ are small for $\beta \lesssim 6.0$
when spiral motion (SP) is dominant for Models D and G,
while they are large 
if buoyant bubbles (BB) are generated
for Models B and E. 
When the intermediate patterns between SP and BB appear in the non-rotational models, the variances have moderate values for Models A and H.
The emergence of buoyant bubbles in the spiral flow for SPB and the pulsating spiral motion for SPP prevents the global rotation of matter in the gain region.
Hence, the centrifugal force, acting on the matter toward perpendicular to the rotation axis, can be reduced. 
Furthermore, for Models D and G, the expanding directions of the shock waves  abruptly become unstable as $\beta$ increases. 
Eventually, the directions, in which the shock waves are revived at first, are not confined neither along the rotation axis nor along the equatorial plane.

\begin{figure}
\epsscale{1.0}
\plotone{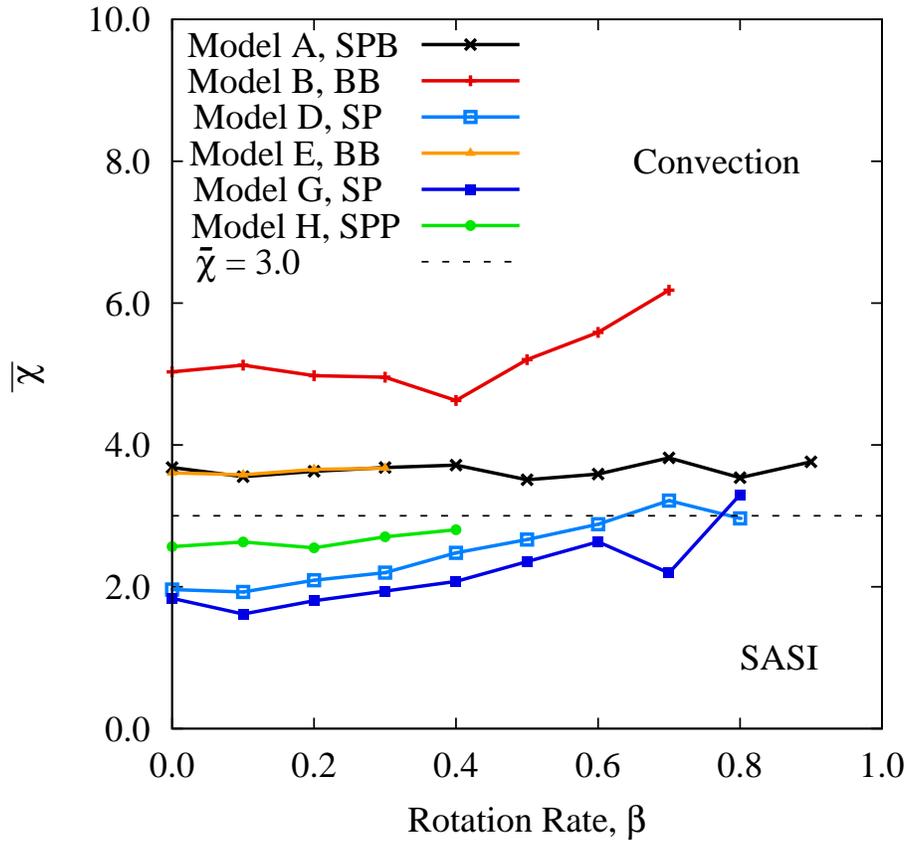}
\caption{
The parameter $\bar{\chi}$ as a function of $\beta$ for non-explosion models.
}
\label{fig_kai}
\end{figure}

Why do the directions, in which the shock waves begin to run away at first, become unstable with increasing the rotation rates for Models D and G?
To answer the question, we present $\bar{\chi}$ as a function of $\beta$ for non-explosion models in Figure~\ref{fig_kai}.
The parameter $\chi$, which is the ratio of the advection time to the growth time of 
 convection,
is a good indicator of the emergence of buoyant bubbles generated by 
neutrino-driven convection,
where its criterion is $\chi>3$ \citep{foglizzo06}.
In this study, we calculate $\bar{\chi}$ for each $\beta$.
$\bar{\chi}$ is obtained from the mean flow averaged over the solid angle during the quasi-steady state, and how to calculate $\bar{\chi}$ are elaborated in Appendix \ref{appex_chi}.
We confirmed that buoyant bubbles emerge if $\bar{\chi}>3$.
It is acceptable that the intermediate patterns have $\bar{\chi}\sim3.5$ for SPB and $\bar{\chi}\sim 2.5$ for SPP,
because in the former pattern buoyant bubbles can be clearly seen in the spiral flow, but not in the latter one. 
Hence, we term the regions above and below $\bar{\chi} = 3$ as 
neutrino-driven convection
and SASI, respectively.
Models D and G exist in the SASI region at $\beta=0$.
However, $\bar{\chi}$ grows with increasing $\beta$, and exceeds the line of $\bar{\chi} = 3$ at $\beta=0.7$ for Model D and at $\beta=0.8$ for Model G.
Then, the position angle $\theta$, at which the shock wave is most extended, becomes unstable (Figure~\ref{fig_th} (d) and (g)).
Here, we consider that the rapid rotation changes the flow dynamics for Model D and G, i.e. the dominant instability is changed from SASI to 
neutrino-driven convection
by rotation.
The rapid rotation can induce the large centrifugal force, which can more effectively act on the flow with spiral motion than on it with buoyant bubbles,
and then the shock radius is enlarged.
The larger the shock radius is, the longer the advection time tend to be.
Hence, the accretion flow might acquire enough time to grow buoyant bubbles,
and then the direction, in which the shock wave is revived at first, becomes unstable.
On the other hand, for Models A, B, H and E,
$\bar{\chi}$ keeps in the same region even if $\beta$ increases.
So the time evolutions of $\theta$ do not show any drastic changes in their characteristics (Figure~\ref{fig_th} (a), (b), (e), and (h)).

\begin{figure}
\epsscale{.55}
\plotone{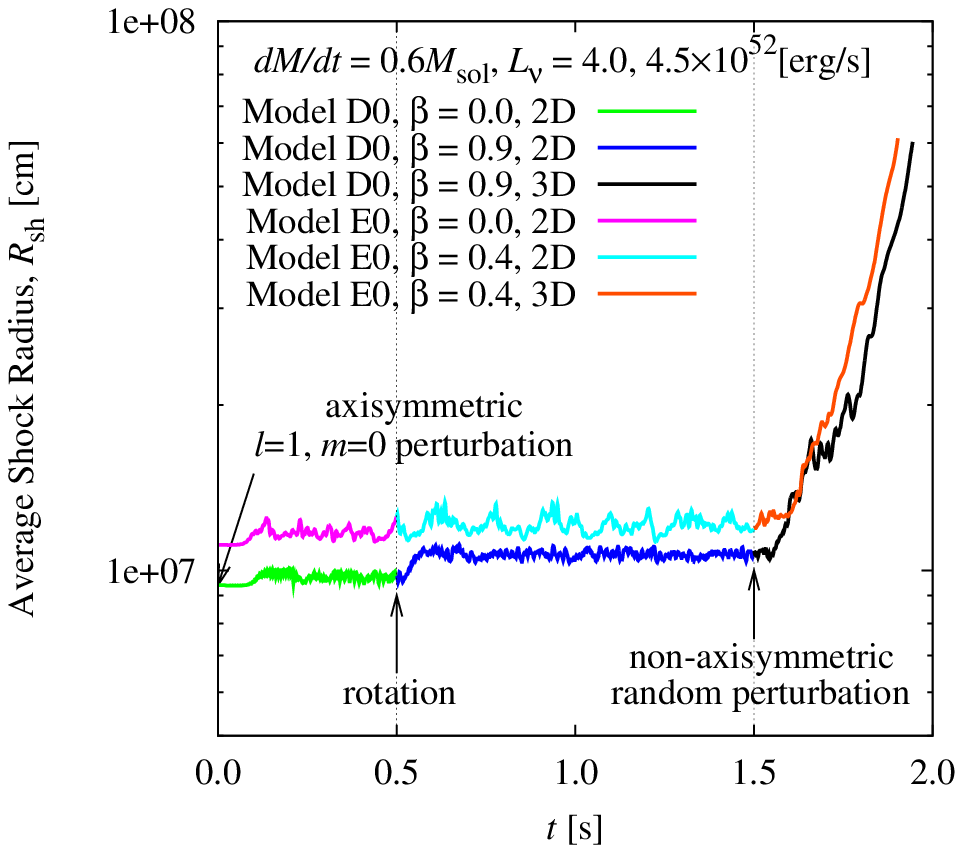}
\plotone{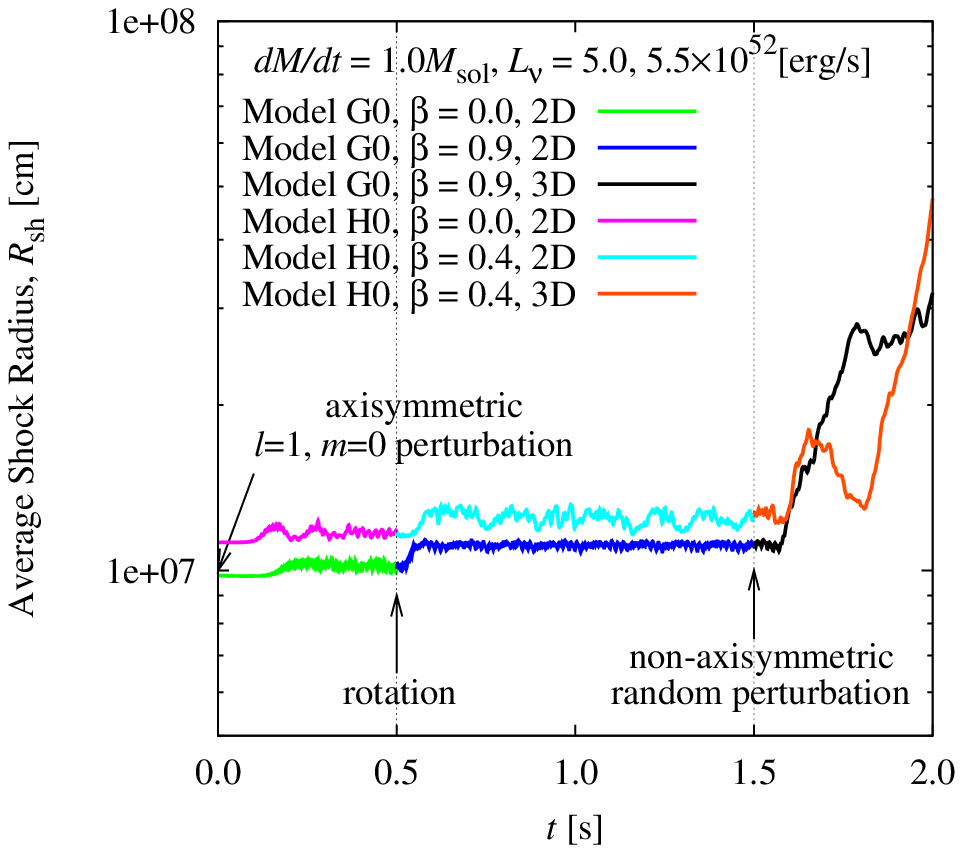}
\caption{
Time evolutions of the averaged shock radius for $\dot{M} = 0.6 M_\odot$ (upper panel) and $\dot{M}=1.0 M_\odot$ (lower panel).
Initially, the $l=1, m=0$ perturbation is imposed on the spherically symmetric flow at $t=0$,
and the 2D models without rotation are obtained from $t=0.0$ to $0.5$.
Next, the rotating matter is injected from the outer boundary into the 2D flows at $t=0.5$,
and the 2D models with rotation is obtained from $t=0.5$ to 1.5.
Finally, the non-axisymmetric random perturbation is added to the 2D flows with rotation, and the 3D rotational models can explode.
\label{fig_2D}}
\end{figure}

Finally, we refer to the difference between 2D and 3D.
The time evolution of the averaged shock radius in 2D and 3D simulations for $\dot{M}=0.6M_\odot$ and 1.0$M_\odot$ is shown in Figure~\ref{fig_2D}.
There are three steps: (1) the 2D models without rotation from $t=0$ to $0.5$ s, (2) the 2D models with rotation from $t=0.5$ to $1.5$ s, and (3) the 3D models with rotation from $t=1.5$ to 2.0 s.   
We find that the explosions occur only in the 3D models at the parameters on the critical surface.
This means that $m\ne0$ modes of 
neutrino-driven convection
or SASI play an important role for rotational explosions at least around the critical surface.
The centrifugal force, acting on the fragmented matter by hydrodynamical instability, further deforms the shock wave and enhances the neutrino heating behind the shock wave. 
It is also confirmed
in more realistic simulations by other groups
that 3D models with rotation is easier to explode than 2D
\citep[][T. Takiwaki, private communication]{nakamura14},
but it has a possibility to depend on the parameters such as a rotation rate.
In order to clarify the whole picture of this topic, more 3D parametric studies at the different parameters are interesting, and its detailed mechanism should be investigated by more realistic simulation with high resolution grids.

\section{SUMMARIES AND DISCUSSIONS}

In this paper, we investigated how rotation affects the shock revival for core-collapse supernovae under the situations in which non-axisymmetric hydrodynamical instabilities are developed.
In order to obtain a critical surface, above which the shock wave can run away, in the three-dimensional parameter space $(\dot{M}, L_\nu, \beta)$,
the rotating matter is injected from the outer boundary into the supersonic flow with increasing $\beta$ every 500 ms, where $\dot{M}$, $L_\nu$, $\beta$ are the mass accretion rate, the neutrino luminosity, and the normalized specific angular momentum, respectively.
Based on the results by \citet{heger05}, the range of the specific angular momentum is taken to be $10^{15}-10^{16}$ cm$^{2}$ s$^{-1}$ in this study.

We found that:
(1) there is a critical value of $\beta$ for a given combination of $\dot{M}$ and $L_\nu$, (2) rotation makes a stalled shock wave easier to revive under the development of the three-dimensional hydrodynamical instabilities, (3) the neutrino heating mechanism works for the range of the specific angular momentum for $10^{15}-10^{16}$ cm$^2$ s$^{-1}$, (4) the direction, in which the shock wave is most extended, tends to be perpendicular to the rotation axis, (5) the dispersion of its direction from the equatorial plane depends on the dominant hydrodynamical instability of SASI and 
neutrino-driven convection,
(6) the rapid rotation can change the dominant instability from SASI to 
neutrino-driven convection,
(7) $m\ne0$ modes play an important role to the rotational explosions.

From (4) and (5), we estimate that the probability distributions of the expanding direction of the shock wave depend on the dominant hydrodynamical instability.
The average direction is perpendicular to the rotation axis,
while their dispersions tend to be small for SASI and large for 
neutrino-driven convection.
In this study, the 
neutrino-driven convection 
is more dominant than SASI at the shock revival due to (6), but the dominant instability might be changed in the different parameters.
Furthermore, if the time scale of flow motion driven by strong rotation is shorter than the growth times of these instabilities, the direction, in which the shock wave is extended at first, might change from perpendicular to the rotation axis to parallel to it.
Hence, the survey of the broader range of the parameters ($L_\nu$, $\dot{M}$, $\beta$) with a large number of realizations is interesting to understand the overall picture of the explosion of the rotating stars. 

In this study we assumed to be a constant mass accretion rate, a constant proto-neutron star mass, and a constant neutrino luminosity.
The initial conditions are the steady solutions with a random perturbation.
Hence, the results should be confirmed in the context of the dynamical evolutions of the actual profiles of the supernova progenitors.
Moreover, the results should be also validated with more accurate treatment of neutrino transfer, which will be addressed by using Boltzmann solver for neutrino radiation hydrodynamics \citep{sumiyoshi14, nagakura14}.
It is also interesting to know the neutrino signal and gravitational wave, emitted from the rotating matter behind the shock wave.
These problems should be investigated in the future work.

\acknowledgments

Numerical computations were performed on the XC30 and the general common use computer system at the center for the Computational Astrophysics, CfCA, the National Astronomical Observatory of Japan, as well as, the Altix UV 1000 at the IFS in Tohoku University and SR16000 at YITP in Kyoto University.
This study was supported by the Grants-in-Aid for the Scientific Research (NoS.~24244036, 24740165), the Grants-in-Aid for the Scientific Research on Innovative Areas, "New Development in Astrophysics through multi messenger observations of gravitational wave sources" (No.~24103006), and the HPCI Strategic Program from the Ministry of Education, Culture, Sports, Science and Technology (MEXT) in Japan.

\appendix

\section{Formulation for Rotational Velocity \label{appex_rot}}

The mass accretion rate is written in
\begin{equation}
\dot{M}(r) \equiv \int_0^{4\pi} \rho v_r r^2 d\Omega,
\end{equation}
and the angular momentum passing through the spherical boundary per unit time is described as
\begin{equation}
\dot{l}_z (r) \equiv \int_0^{4\pi} \rho v_r (v_\phi r \sin\theta) r^2 d\Omega,
\end{equation}
where $\rho$, $v_r$, $v_\theta$, and $\Omega$ are the density, radial velocity, azimuthal velocity, and solid angle, respectively. 
In the rigid rotation, the azimuthal velocity is defined as
\begin{equation}
v_\phi (r,\theta) = r \sin \theta \  \omega_z (r),
\label{eq_vph2}
\end{equation}
where $\omega_z$ is the angular velocity of rotation around the polar axis.
If we assume the steady and spherically symmetric distributions of density and radial velocity in the supersonic flow above the shock wave, the angular momentum rate can be written in
\begin{equation}
\dot{l}_z  = \frac{2}{3} \dot{M} r^2 \omega_z (r),
\end{equation}
where $\dot{l}_z$ and $\dot{M}$ are constant along the radial direction
for the mass and angular momentum conservation.
Hence, we can obtain
\begin{equation}
r^2 \omega_z (r) = C,
\label{eq_const}
\end{equation}
where $C$ is a constant.
We substitute Eq.~(\ref{eq_const}) into Eq.~(\ref{eq_vph2}), and the azimuthal velocity can be described as
\begin{equation}
v_\phi (r,\theta) = \frac{C}{r} \sin \theta.
\label{eq_vph3}
\end{equation}
We set the constant to be $C = \beta\times10^{16}$,
that is, $\omega_z=\beta$ rad s$^{-1}$ at $r=$1000 km.

\begin{figure}[h]
\epsscale{1.00}
\plotone{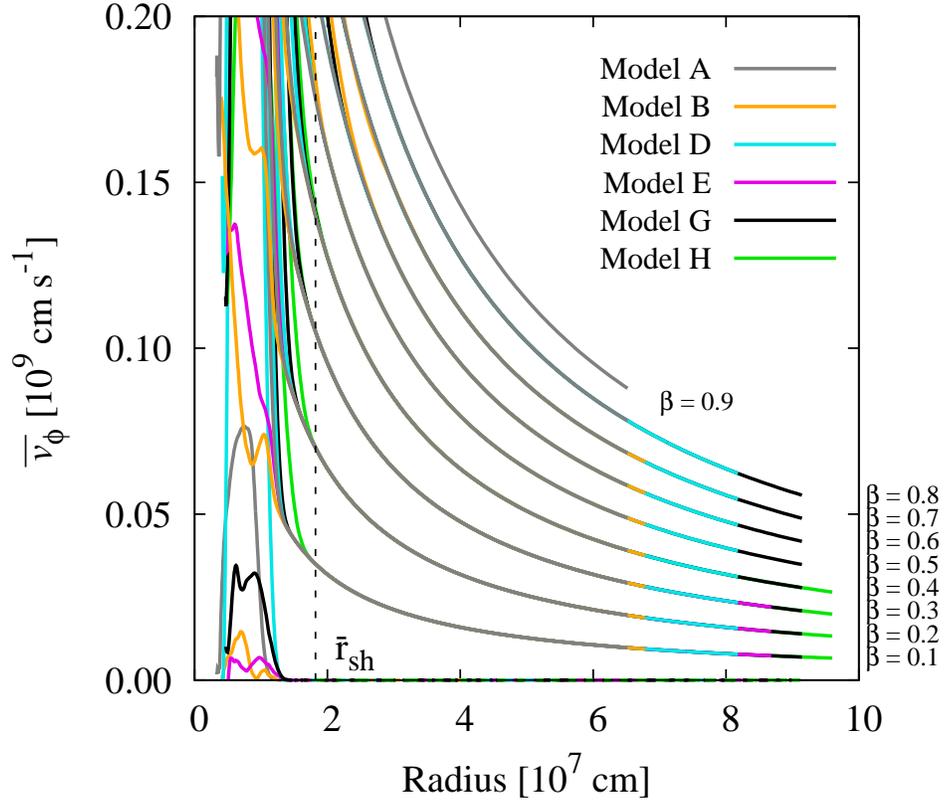}
\caption{
The radial distributions of the time- and angle-averaged azimuthal velocity $\bar{v}_\phi$ for the non-explosion models.
The dashed line at $\bar{r}_{\rm sh}$ means the maximum radius of the shock wave of all models.
The locations of the right endpoints of the lines correspond to the outer boundaries, respectively.
\label{fig_vph}}
\end{figure}

In order to investigate the validity of the assumptions described above, the radial distributions of the time- and angle-averaged azimuthal velocity $\bar{v}_\phi$ for Models A, B, D, E, G, and H are shown in Fig.~\ref{fig_vph}.
The right endpoints of the lines correspond to the outer boundaries, and the dashed line at $\bar{r}_{sh}$ indicates the maximum shock radius of all models.
The lines having same $\beta$ are overlapped each other for $\bar{r}_{sh} \ll r$, although the lines are deviated from one another for $r \lesssim \bar{r}_{sh}$.
Therefore, the assumption is valid for $\bar{r}_{sh} \ll r$, and the angular velocities are appropriately given to the outer boundaries even if the radii of the outer boundaries are different among models.

\section{Average Shock Radius and Mode Analysis \label{appex_sph}}

The distance of the shock front from the origin can be written in a linear combination of the spherical harmonic components,
\begin{equation}
R_\mathrm{sh}(\theta, \phi, t) = \sum^{\infty}_{l=0} \sum^{l}_{m=-l} c^m_{l} (t) \, Y^m_{l}(\theta, \phi),
\label{eq_sph}
\end{equation}
where $Y^m_{l}$ is expressed by the associated Legendre polynomial $P^m_{l}$ as
\begin{equation}
Y^m_{l} = K^m_{l} P^m_{l}(\cos \theta) \, e^{im\phi},
\label{eq_y}
\end{equation}
\begin{equation}
K^m_{l} = \sqrt{\frac{2l+1}{4\pi}\frac{(l-m)!}{(l+m)!}}.
\label{eq_k}
\end{equation}
The expansion coefficients are described as
\begin{equation}
c^m_{l} (t) =\int^{2\pi}_0 \! \! \! \! d\phi  \! \int^{\pi}_0 \! \! d\theta \, \sin \theta \, R_\mathrm{sh}(\theta, \phi, t) \, Y^{m*}_{l} (\theta, \phi),
\label{eq_c}
\end{equation}
where the superscript * denotes complex conjugation.

The following quantities:
\begin{equation}
A_l(t) = \sqrt{\Sigma^l_{m=-l} |c^m_l(t)/c^0_0(t)|^2},
\label{eq_A}
\end{equation}
\begin{equation}
A_{1, 2}(t) =  \sqrt{\Sigma^2_{l=1}\Sigma^l_{m=-l} |c^m_l(t)/c^0_0(t)|^2},
\label{eq_A12}
\end{equation}
\begin{equation}
A_{4, 5} (t) =  \sqrt{\Sigma^5_{l=4}\Sigma^l_{m=-l} |c^m_l(t)/c^0_0(t)|^2},
\label{eq_A45}
\end{equation}
and their time-averages:
\begin{equation}
\bar{A}_l = \frac{1}{T}\int^{t_e}_{t_s} A_l(t) dt,
\label{eq_Ab}
\end{equation}
\begin{equation}
\bar{A}_{1,2} = \frac{1}{T}\int^{t_e}_{t_s} A_{1,2}(t) dt,
\label{eq_A12b}
\end{equation}
\begin{equation}
\bar{A}_{4,5} = \frac{1}{T}\int^{t_e}_{t_s} A_{4,5}(t) dt,
\label{eq_A45b}
\end{equation}
are used for the mode analysis, 
where $T=t_e-t_s$ is the integral time, $t_s$ is the starting time, and $t_e$ is the ending time for integration.

\section{Parameter $\chi$ \label{appex_chi}}

Taking into account the effect of advection, \citet{foglizzo06}
proposed the parameter $\chi$ as the new criterion of convective instability, 
which is the ratio of advection time to growth time of buoyancy, is defined as
\begin{equation}
\chi \equiv\int_{r_{\rm gain}}^{r_{\rm sh}} \left|\frac{N}{v_r}\right| dr,
\label{eq_chi}
\end{equation}
where
$r_{\rm gain}$ is the gain radius, $r_{\rm sh}$ is the shock radius,
and $v_r$ is the radial velocity.
The Brunt-V\"{a}is\"{a}l\"{a} frequency $N$ can be written in
\begin{equation}
N^2 =\left|\frac{1}{\Gamma_1 p}\frac{dp}{dr}-\frac{1}{\rho}\frac{d\rho}{dr}\right|g,
\end{equation}
\begin{equation}
\Gamma_1 = \left(\frac{\partial \ln{p}}{\partial \ln{\rho}}\right)_{S, Y_e},
\label{eq_gamma}
\end{equation}
where $p$, $\rho$, $S$, $Y_e$, and $g$ are the pressure, density, entropy, electron fraction, and gravitational acceleration, respectively.
The gravitational acceleration is given approximately in the gain region as $g = \frac{GM_{\rm PNS}}{r^2}$, in which $M_{\rm PNS}$ and $G$ is the PNS mass and the gravitational constant.

Furthermore, for three-dimensional quasi-steady models, we define the angle-averaged mean flow, where the quantities are averaged over the solid angle and from the onset of the quasi-steady state to the end of computation,
\begin{equation}
\bar{q}_{\rm 1D}(r) = \frac{1}{4\pi}\int_0^{4\pi} \left[\frac{1}{T}\int^{t_e}_{t_s} q(r,\theta,\phi,t) dt \right] d\Omega.
\label{eq_ab}
\end{equation}
where $q(r,\theta,\phi,t)$ is an arbitrary quantity (i.e., $\rho$, $v_r$, and so on), $t_s$ is the beginning time of the quasi-steady state, and $t_e$ is the ending time of the computation.
The parameter $\bar{\chi}_{\rm 1D}$ is then obtained as
\begin{equation}
\bar{\chi}_{\rm 1D} = \int_{\bar{r}_{\rm gain 1D}}^{\bar{r}_{\rm sh 1D}} \left|\frac{\bar{N}_{\rm 1D}(r)}{\bar{u}_{r {\rm 1D}}(r)}\right| dr,
\label{eq_kaib}
\end{equation}
\begin{equation}
\bar{N}^2_{\rm 1D} =\left|\frac{1}{\bar{\Gamma}_{1\rm 1D} \bar{p}_{\rm 1D}}\frac{d\bar{p}_{\rm 1D}}{dr}-\frac{1}{\bar{\rho}_{\rm 1D}}\frac{d\bar{\rho}_{\rm 1D}}{dr}\right|g,
\label{eq_nb}
\end{equation}
where the quantities with a bar involves the angle-averaged mean flow.
The thermodynamical variables, $\bar{p}_{\rm 1D}$ and $\bar{\Gamma}_{1\rm 1D}$, are given by an EOS table as a function of $\bar{\rho}_{\rm 1D}$, $\bar{e}_{\rm 1D}$ and $\bar{Y}_{e\rm 1D}$.
$\bar{r}_{\rm gain 1D}$
is the gain radius where the net heating rate of the angle-averaged mean flow is equal to zero, and 
$\bar{r}_{\rm sh 1D}$ is
the shock radius where $\bar{S}$ is equal to 3.1.

\section{Model B \label{appex_B}}

In this appendix, we summarize the characteristics of Model B
for $\dot{M}=0.2M_\odot$, $L_\nu=2.75\times10^{52}$ erg s$^{-1}$, and $\beta=0.0$.
Flow patterns in semi-nonlinear/nonlinear phases and physical parameters are listed in Table \ref{tbl_modelB}.
The time evolutions of normalized mode amplitudes, the orientations of the rotation axis, and the magnitudes of angular momentum are shown in Figure \ref{fig_modelB}.
Mode amplitude $A_l$ and parameter $\chi$ are defined in Appendix \ref{appex_sph} and \ref{appex_chi}, respectively.
Model B have two patterns in the semi-nonlinear phase.
Depending on the initial perturbations which is made from the different seed of random number, 
either the formation of a spiral flow (SP) or the emergence of buoyant bubbles (BB) can be seen in the semi-nonlinear phase
(Figure~\ref{fig_modelB} (a), (b)).
The oscillating growth of lower modes corresponds to the growth of spiral motion of the shock wave,
while the monotonic increase of various modes indicates the shock deformation by rising buoyant bubbles.
Hence, the parameter $\chi=4.1$, which is obtained from the initial flow for Model B, might be a critical value in our simulations.
In the nonlinear phase, the flow pattern turns into BB for both Models B1 and B2, judged 
from the unstable behavior of the rotation axis (Figure~\ref{fig_modelB} (d), (e)) and the low angular momentum (Figure~\ref{fig_modelB} (g), (h)),
where $\bar{\chi}$ is about 5.0.
Furthermore, we have done the resolution test for Model B.
Flow patterns in the high resolution models are consistent with ones in the normal resolution models.
In the semi-nonlinear phase, BB emerges in the middle resolution model (left panel in Fig.~\ref{fig_modelB}(c)), and SP appears in the high resolution model (right panel in Fig.~\ref{fig_modelB}(c)).
In the nonlinear phase, BB can be seen in both middle and high resolution models.
Although the ratio of mode amplitudes $\bar{A}_{l=1,2}/\bar{A}_{l=4,5}$ decrease with increasing resolution,
the difference of resolution does not impact on flow patterns \citep[see][]{iwakami14}.

\section{Grid Resolution \label{appex_res}}

\begin{figure}
\epsscale{0.8}
\plotone{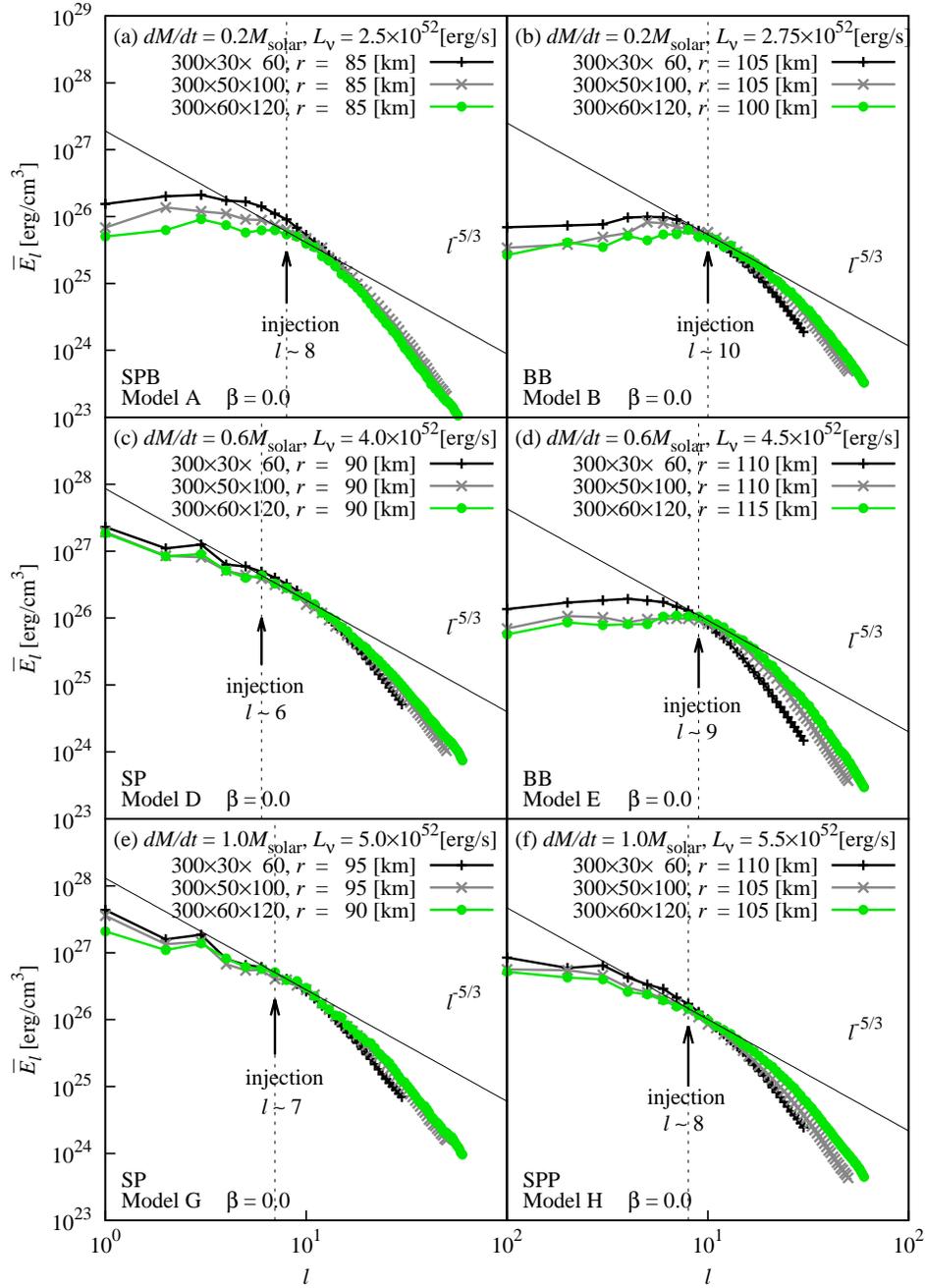}
\caption{
The time-averaged kinetic-energy spectra $\bar{E}_l$ of turbulence as a function of the polar index $l$ for non-rotational models.
The results for three different resolutions, $300\times 30\times 60$ (black lines), $300\times 50 \times 100$ (gray lines), and $300\times 60\times 120$ (green lines) grid points are compared.
The radial positions where we evaluate $\bar{E}_l$ correspond to the points, at which the maximum values of $|\bar{N}_{1D}/\bar{v}_{r 1D}|$ are attained. Here $\bar{N}_{1D}$ denotes the Brunt-V\"{a}is\"{a}l\"{a} frequency and $\bar{v}_{r 1D}$ is the radial velocity given by Equation~(\ref{eq_kaib}).
According to the results for $300\times60\times120$ grid points,
it is inferred that the energy injections occur around $l\sim 6-10$.
}
\label{fig_turb}
\end{figure}

In this appendix we study whether the spatial resolution employed in this paper is good enough.
For that purpose we compare simulations of both rotational and non-rotational models with different numbers of grid points.
We change only angular resolutions
because the radial grid spacings are already $\sim 10$ 
times finer than the angular ones and a further increase in the radial resolution will
hardly affect the outcome.
We first compare non-rotational models computed with $N_r \times N_\theta \times N_\phi = 300\times30\times60$ (fiducial), $300\times50\times100$ and $300\times60\times120$ mesh points.
For none of these models we obtain shock revival.
We then utilize the quasi-steady turbulences obtained at the end of the simulations for comparison.
We present the kinetic-energy spectra of turbulence in Figure \ref{fig_turb}.
Following \citet{hanke12, dolence13, couch13a}, we define the energy spectra $E_l$ as
\begin{equation}
E_l(t,r) = \sum^l_{m=-l} \left|\int_\Omega Y_l^{m *}(\theta, \phi) \sqrt{\rho(v_\theta^2 + v_\phi^2)} d\Omega \right|^2,
\end{equation}
where $Y_l^{m *}$ is the complex conjugation of the spherical harmonics of order $(l,m)$ given in Equation (\ref{eq_y}), $\Omega$ is the solid angle, $\rho$ is the density, and $v_\theta$ and $v_\phi$ are the polar and azimuthal components of velocity, respectively.
In the following, we use the spectra $\bar{E}_l(r)$ averaged over the quasi-steady phase as
\begin{equation}
\bar{E}_l (r) = \frac{1}{T}\int^{t_e}_{t_s} E_l(t,r) dt,
\end{equation}
where $T=t_e-t_s$ is the integral time, with $t_s$ and $t_e$ being the start and end times of integration.
$\bar{E}_l$ is evaluated at the radial point, at which the local maximum value of $|\bar{N}_{1D}(r)/\bar{v}_{r 1D}(r)|$ is attained, where $\bar{N}_{1D}$ and $\bar{v}_{r 1D}$ denote the Brunt-V\"{a}is\"{a}l\"{a} frequency and the radial velocity, respectively, defined in Appendix \ref{appex_chi}.

In the Figure \ref{fig_turb}, the injection scale of the turbulent energy is investigated, corresponding to the lowest $l$ in the inertial range in which $\bar{E}_l$ is proportional to $l^{-5/3}$.
From the results obtained by using the finest meshes, we infer that the 
turbulent energy is injected around $l\sim 6-10$, which is roughly same as the findings in some papers.
\citet{hanke12} and \citet{couch13a} suggested that the injection scale is $l\sim 10$, and
\citet{dolence13} found that the scale is $l\sim4$.
We can see that the
values of $\bar{E}_l$ around injection scale
agree well with each other
among simulations with the different resolutions.
At larger scales than their injection scales, they are still not very different qualitatively although the turbulent energy is slightly overestimated in the fiducial cases.
In fact, 
the spectral peaks are obtained at
$l=1$ or 2 for Models D, G, and H (Fig.~\ref{fig_turb} (c), (e), and (f)),
while they occur at $3 \le l < 10$ for Models A, B, and E (Fig.~\ref{fig_turb} (a), (b), and (d)).
They reflect the fact that in the former models, 
coherent spiral and/or sloshing motions occur 
whereas buoyant bubbles are formed copiously
in the latter, which is unaffected by the difference of resolution.
As mentioned in the main body, 
SASI is thought to be in operation in the former case
and the neutrino-driven convection is supposed to be dominant in the latter.
We repeat, however, that the important point here is the fact that the large scale flow pattern, which is crucial for the enhancement of neutrino heating, is not changed by the increase of grid points.
In this sense we can say that the grid we employed in this paper is good enough.

It is admittedly true that the grid resolution is not sufficient to obtain the inertial range unambiguously.
In the previous papers \citep{hanke12, couch13a},
the kinetic-energy spectrum obeys the Kolmogorov's law, $\propto l^{-5/3}$, in the inertial range.
\citet{dolence13}, however, found that the spectrum has the power-law slope of -1.
Recently, \citet{couch13b} indicated by the high-resolution simulations that the spectra are consistent with a -1 power-law for $10\lesssim l \lesssim 40$ and a -5/3 power-law for $40 \lesssim l \lesssim 80$.
Hence, they considered that the injection point might be around $l\sim40$.
In Figure \ref{fig_turb} such features are barely seen with the higher resolutions.
This is due to numerical viscosities, i.e. smaller-scale $(l>10)$ features are smeared out.
It was reported by \citet{murphy13} that the ram pressure of turbulent motions generated by the neutrino-driven convection may be large enough to enlarge a shock front.
This is indeed the case, we may have overestimated the critical luminosity although the overestimation of turbulence power in large scales may somewhat compensate for this effect.

\begin{figure}
\epsscale{1.0}
\plotone{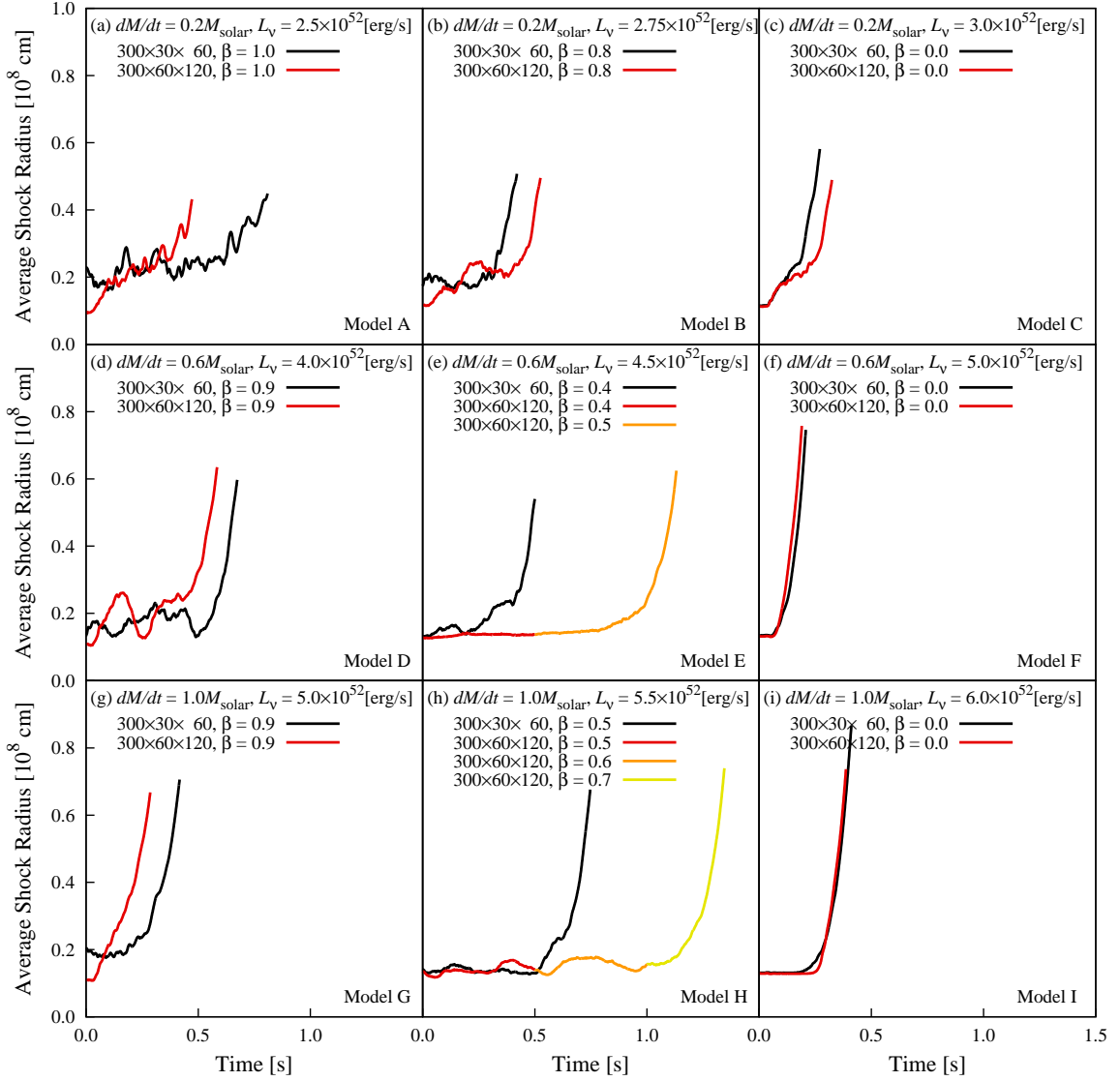}
\caption{
The evolutions of the averaged shock radius. 
The results for the fiducial resolution of $300\times 30\times 60$ grid points (black lines) are compared with those for the higher resolution of $300 \times 60 \times 120$ grid points (colored lines).
}
\label{fig_rshg}
\end{figure}

Next, we examine the rotational models with a higher resolution with $300\times60\times120$ grid points.
We are concerned here with to what extent $\beta_\mathrm{exp}$ is affected.
Considering limited numerical resources,
we inject the rotational flow with $\beta$=$\beta_\mathrm{exp}$ from the beginning for the higher resolution.
If it does not lead to shock revival within 500ms, $\beta$ is increase at that point.
Figure\ref{fig_rshg} shows the evolutions of shock radius.
In most of the cases, the shock-revival time is changed.
It can be earlier or later, reflecting the stochastic nature of shock revival \citep{nagakura13, takiwaki14}.
On the other hand, $\beta_\mathrm{exp}$ is unchanged for all but two models.
We find that $\beta_\mathrm{exp}$ is modified to a bit higher values for Models E and H.
It should be emphasized that the main conclusion that rotation lowers the critical luminosity is intact.

\begin{figure}
\epsscale{1.0}
\plotone{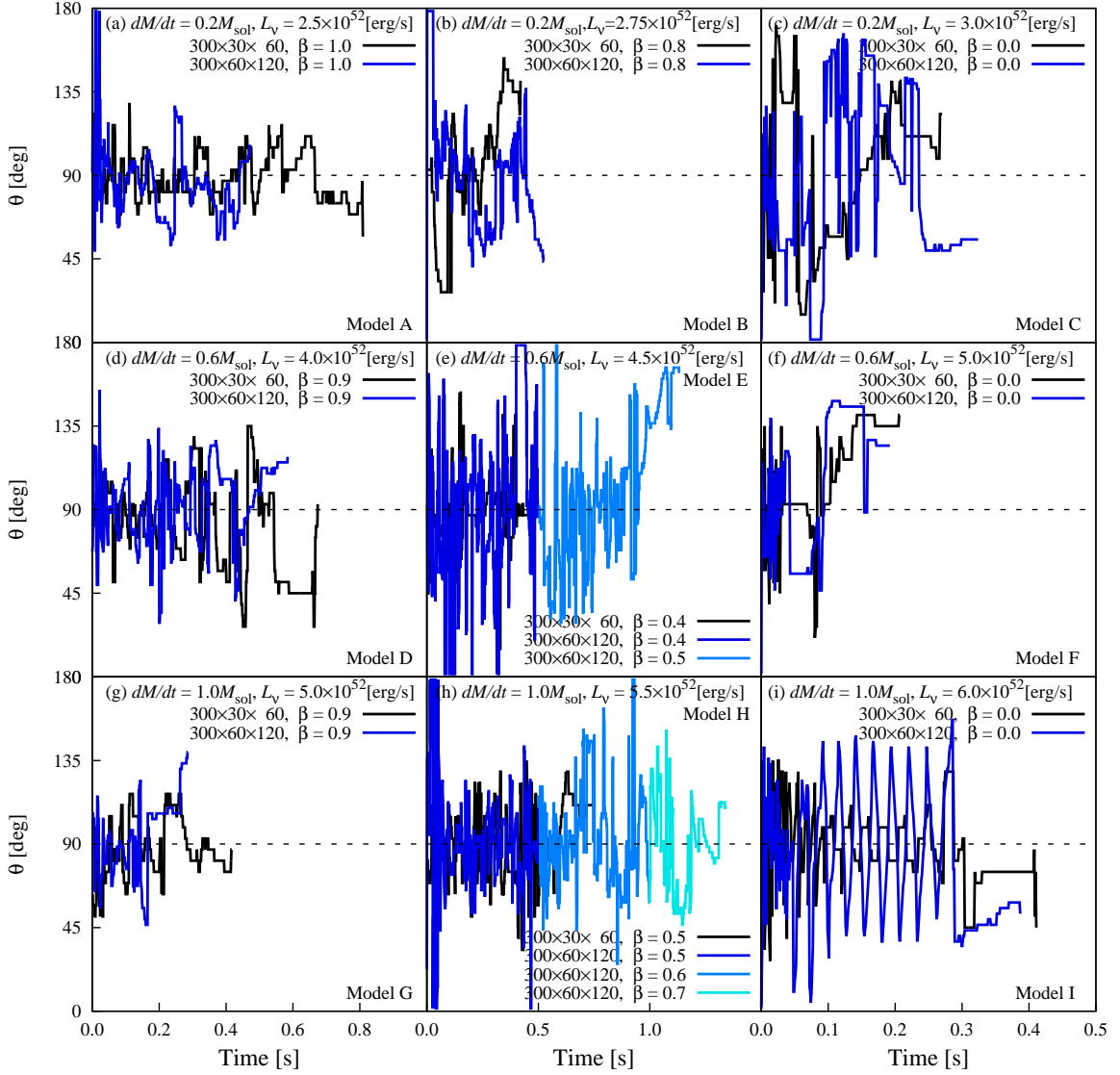}
\caption{
The evolutions of the position angle $\theta$, at which the shock wave is most extended. 
The results for the fiducial resolution of $300\times 30\times 60$ grid points (black lines) are compared with those for the higher resolution of $300 \times 60 \times 120$ grid points (colored lines).
}
\label{fig_thg}
\end{figure}

Figure~\ref{fig_thg} shows the evolutions of the 
position angle, at which the shock is most expanded.
It can be seen from this figure that the shock revival tends to occur perpendicularly to the rotation axis regardless of resolution.
Moreover, the average deviations of $\theta$ from 90 deg are more or less the same between normal- and high-resolution models.
We hence believe that the fiducial resolution of $300\times30\times60$ is justified also for rotational models. 

\clearpage

\begin{deluxetable}{cccccccc}
\tablecaption{Summary of parameters for all models.\label{tbl_param}}
\tablewidth{14cm}
\tablehead{
Model &
$\dot{M}$\tablenotemark{a} &
$L_{\nu}$\tablenotemark{b} &
$r_\mathrm{in}$\tablenotemark{c} &
$r_\mathrm{out}$\tablenotemark{d} &
Flow Patterns &
$\beta_\mathrm{exp}$\tablenotemark{e}\\
& 
[$M_\odot$ $s^{-1}$] &
[$10^{52}$ erg $s^{-1}$] &
[km] &
[km] &
(Nonlinear, $\beta=0$) &
}
\startdata
A & 0.2 & 2.5  & 33 & 655  & SPB & 1.0     \\
B & 0.2 & 2.75 & 34 & 688  & BB  & 0.8     \\
C & 0.2 & 3.0  & 36 & 712  & BB  & 0.0     \\
D & 0.6 & 4.0  & 41 & 822  & SP  & 0.9     \\
E & 0.6 & 4.5  & 44 & 872  & BB  & 0.4     \\
F & 0.6 & 5.0  & 46 & 920  & BB  & 0.0     \\
G & 1.0 & 5.0  & 46 & 919  & SP  & 0.9     \\
H & 1.0 & 5.5  & 49 & 965  & SPP & 0.5     \\
I & 1.0 & 6.0  & 51 & 1007 & SP  & 0.0     \\
\enddata
\tablenotetext{a}{Mass accretion rate.\vspace{-5mm}}
\tablenotetext{b}{Neutrino luminosity.\vspace{-5mm}}
\tablenotetext{c}{Radius of the inner boundary.\vspace{-5mm}}
\tablenotetext{d}{Radius of the outer boundary.\vspace{-5mm}}
\tablenotetext{e}{Critical value of the normalized specific angular momentum.\vspace{-5mm}}
\tablecomments{
Flow patterns are classified into sloshing motion (SL), spiral motion(SP), buoyant bubbles (BB), spiral motion with rising buoyant bubbles (SPB), and spiral motion with pulsating rotational velocity (SPP).
}
\end{deluxetable}

\clearpage

\begin{deluxetable}{cccccccccc}
\tablecaption{Flow patterns and physical parameters for models B.\label{tbl_modelB}}
\rotate
\tablewidth{21cm}
\tablehead{
Model &
Semi-Nonlinear &
$\chi$\tablenotemark{a} &
$r_{\rm gain}$\tablenotemark{b} &
$r_{\rm sh}$\tablenotemark{c} &
Nonlinear &
$\bar{\chi}$\tablenotemark{d} &
$\bar{r}_{\rm gain}$\tablenotemark{e} &
$\bar{r}_{\rm sh}$\tablenotemark{f} &
$\bar{A}_{l=1,2}/\bar{A}_{l=4,5}$\tablenotemark{g} \\

&
pattern &
(initial flow) &
[km] &
[km] &
pattern &
(nonlinear flow) &
[km] &
[km] &
}
\startdata
B0-3 & SP/BB & 4.1 & 60 & 99 & BB & 5.0-5.1 & 57 & 133-136 & 2.2-2.4 \\
B4   &    BB & 4.1 & 60 & 99 & BB & 5.1     & 56 & 139     & 1.7     \\
B5   &    SP & 4.1 & 60 & 99 & BB & 5.0     & 56 & 132     & 1.5     
\enddata
\tablenotetext{a}{Parameter $\chi$ in the initial flow\vspace{-5mm}}
\tablenotetext{b}{Radius of the boundary between the heating and cooling regions in the initial flow\vspace{-5mm}}
\tablenotetext{c}{Radius of the shock wave in the initial flow\vspace{-5mm}}
\tablenotetext{d}{Parameter $\chi$ in the time- and angle-averaged flow\vspace{-5mm}}
\tablenotetext{e}{Radius of the boundary between the heating and cooling regions in the time- and angle-averaged flow\vspace{-5mm}}
\tablenotetext{f}{Radius of the shock wave in the time- and angle-averaged flow\vspace{-5mm}}
\tablenotetext{g}{Ratio of the time-averaged mode amplitudes of $A_{l=1,2}$ to $A_{l=4,5}$\vspace{-5mm}}
\tablecomments{
Models B0-3, B4, and B5 correspond to the normal, middle, and high resolution models.
The grid points of normal, middle, and high resolution models are $N_r\times N_\theta \times N_\phi = 300 \times 30 \times 60$, $300 \times 50 \times 100$, and $300 \times 60 \times 120$, respectively.
}
\end{deluxetable}

\clearpage

\clearpage

\clearpage

\begin{figure}
\epsscale{1.0}
\plotone{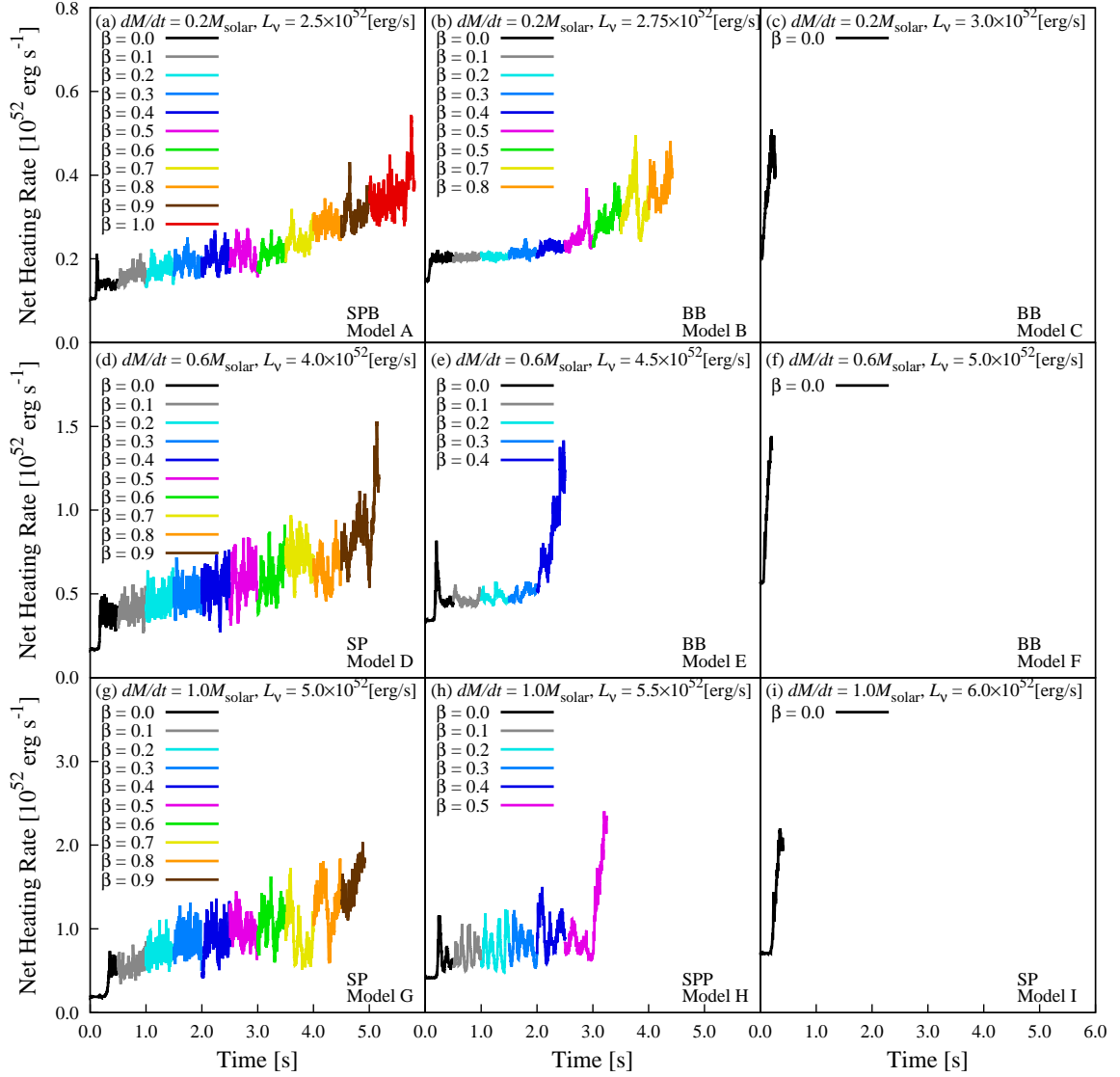}
\caption{
The evolutions of the net heating rate with increasing $\beta$ every 500 ms.
The net heating rate is integrated in the shocked region.
}
\label{fig_nht}
\end{figure}

\clearpage

\clearpage

\clearpage

\clearpage

\clearpage

\clearpage

\begin{figure}
\epsscale{0.90}
\plotone{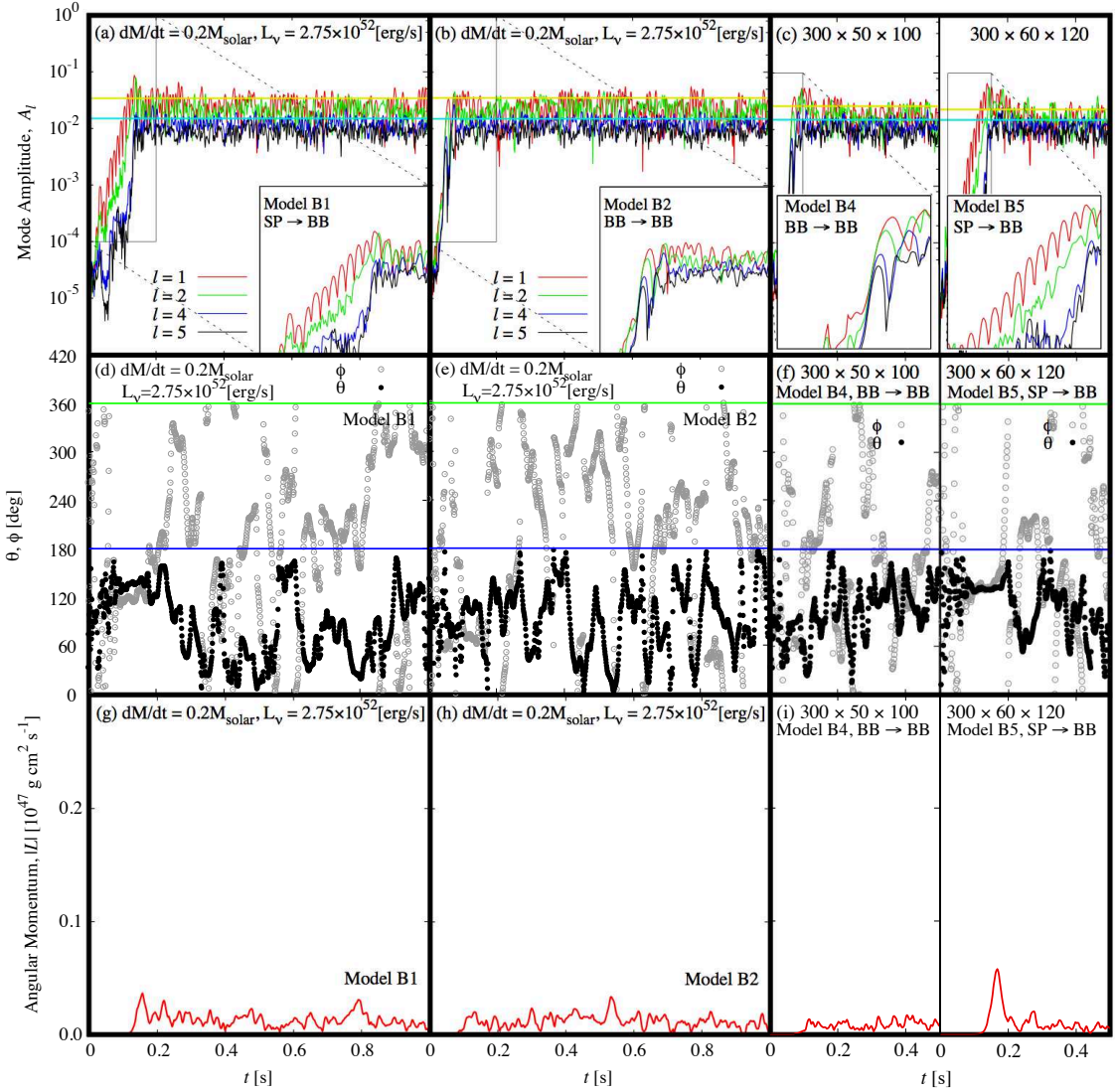}
\caption{
The evolutions of the normalized mode amplitudes, the orientations of the rotation axis, and the magnitudes of the angular momentum for Models B1, B2, B4 and B5. 
In the figures (c), (f) and (i), the left and right panels show the results of 30$\times$50$\times$100 and 30$\times$60$\times$120 mesh points, respectively.
In the figures (a), (b) and (c), the yellow lines correspond to the combined mode amplitudes of $l=1,2$ averaged in the nonlinear phase, whereas the blue lines stand for those of $l=4,5$,
and the insets are the zoom-ups of the indicated portions.
In the figures (d), (e) and (f), the angles $\theta$ (black filled-circle) and $\phi$ (gray open-circle) indicating the orientation of the rotation axis, are the spherical coordinates, 
and the blue and green lines correspond to $\pi$ and 2$\pi$, respectively.
\label{fig_modelB}}
\end{figure}

\clearpage

\clearpage

\clearpage

\end{document}